\documentclass[%
 reprint,
superscriptaddress,
 amsmath,amssymb,
 aps, prl
]{revtex4-1}

\usepackage{graphicx}
\usepackage{dcolumn}
\usepackage{bm}

\begin{document}

\preprint{APS/123-QED}

\title{p-State Luminescence in CdSe Nanoplatelets: \\The Role of Lateral Confinement and an LO-Phonon Bottleneck}

\author{Alexander W. Achtstein${^\dagger}$}
\email{alexander.achtstein@tu-berlin.de}
\altaffiliation[Current address: ]{Optoelectronic Materials Section, Delft University of Technology, 2628 BL Delft, The Netherlands}
\author{Riccardo Scott} \thanks{R.S. and A.W.A. contributed equally to this work.}
\author{Sebastian Kickh\"ofel}
\affiliation{Institute of Optics and Atomic Physics, Technical University of Berlin, Strasse des 17. Juni 135, 10623 Berlin, Germany}
\author{Stefan T. Jagsch}
\affiliation{Institute of Solid State Physics, Technical University of Berlin, Strasse des 17. Juni 135, 10623 Berlin, Germany}
\author{Sotirios Christodoulou}
\affiliation{Department of Physics, University of Genoa, via Dodecaneso 33, IT-16146 Genova, Italy}\affiliation{Istituto Italiano di Tecnologia, Via Morego 30, IT-16163 Genova, Italy}
\author{Anatol V. Prudnikau}
\author{Artsiom Antanovich}
\author{Mikhail Artemyev}
\affiliation{Institute for Physico-Chemical Problems, Belarusian State University, 220030 Minsk, Belarus}
\author{Iwan Moreels}
\affiliation{Istituto Italiano di Tecnologia, Via Morego 30, IT-16163 Genova, Italy}
\author{Andrei Schliwa}
\affiliation{Institute of Solid State Physics, Technical University of Berlin, Strasse des 17. Juni 135, 10623 Berlin, Germany}
\author{Ulrike Woggon}
\affiliation{Institute of Optics and Atomic Physics, Technical University of Berlin, Strasse des 17. Juni 135, 10623 Berlin, Germany}

\date{\today}

\begin{abstract}
We report excited state emission from p-states at excitation fluences well below ground state saturation in CdSe nanoplatelets. Size dependent exciton ground state-excited state energies and dynamics are determined by three independent methods, time-resolved photoluminescence (PL), time-integrated PL and Hartree renormalized \textbf{k$\cdot$p} calculations -- all in very good agreement.  The ground state-excited state energy spacing strongly increases with the lateral platelet quantization. Our results suggest that the PL decay of CdSe platelets is governed by an LO-phonon bottleneck, related to the reported low exciton phonon coupling in CdSe platelets and only observable due to the very large oscillator strength and energy spacing of both states. 

\begin{description}
\item[PACS numbers]
\verb+\pacs{62.23.Kn, 81.07.St, 78.47.da, 73.22.-f, 78.67.Bf}+
\end{description}
\end{abstract}

\pacs{Valid PACS appear here}
\maketitle


Semiconductor nanoparticles have attracted growing attention in the last decade due to their promising optical and electronic properties .  Two dimensional II-VI semiconductor nanoplatelets (NPLs) gained increasing interest due to their unique electronic and optical properties \cite{Antan2015}, such as the Giant Oscillator Strength effect \cite{Ithurria2011a,Achtstein2012,Naeem2015}, room temperature exciton coherence \cite{Cassette2015} and lasing \cite{Grim2014}, strong electroabsorption response \cite{Achtstein2014} and size dependent dark-bright splitting \cite{Biadala2014}. With this letter we report on energies and dynamics of excited state emission from p-states in CdSe nanoplatelets by the means of temperature and time-resolved photoluminescence (PL) and Hartree renormalized \textbf{k$\cdot$p} modeling. Figure 1 shows the evolution of the lowest exciton s and p-states with increasing transversal confinement and anisotropy in CdSe NPLs. 
\begin{figure}[b]
\includegraphics{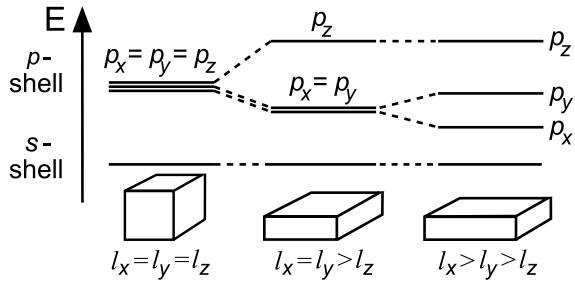}
\caption{Evolution of the electron $p$-shell degeneracy when an isotropic quantum box evolves into a platelet with an unequal side-length.}
\end{figure}

We show in this letter that the ground state-excited state energy difference of CdSe NPLs strongly increases with the lateral platelet quantization and that the already described bi-exponential PL decay of NPLs is connected to the very large dipole moments of the excited state (ES) and ground state (GS) and a phonon bottleneck suppressing inter-relaxation. A dynamic thermal equilibrium between both states mediated by LO phonon scattering is observed. The strong transversal confinement related suppression of LO phonon modes in the NPLs \cite{Achtstein2012}, also observed as lifetime limited dephasing rates \cite{Naeem2015}, results here in a slowdown of the ES--GS exciton transfer rate and subsequent visible ES luminescence well below ground state saturation. We show that the high energy emission originates systematically from the NPL's ES and correlates to the lateral confinement dependent ES--GS energy difference. 
Colloidal 4.5 monolayer (ML) CdSe NPLs were synthesized, characterized by TEM and embedded in PMAO films on thin fused silica substrates mounted in a liquid He cryostat according to detailed description in the Supporting Material.
Our experimental setup allows the consecutive measurement of time-integrated (CCD-spectrometer) and time-resolved (streak camera) fluorescence of a sample with confocal excitation in the platelet absorption continuum ($420$\,nm, SHG of a 75.4\,MHz/150\,fs Ti:Sa laser) and detection  through an (\text{NA}=0.4) objective. The excitation density  was held below moderate 0.2\,W/cm$^2$ to avoid any heating, saturation and the presence of biexcitons (we estimate only $< 0.1$ percent of the platelets are excited within one laser pulse using ICP and absorption cross sections of She et al. \cite{She2014}).
\begin{figure}[t]
\includegraphics{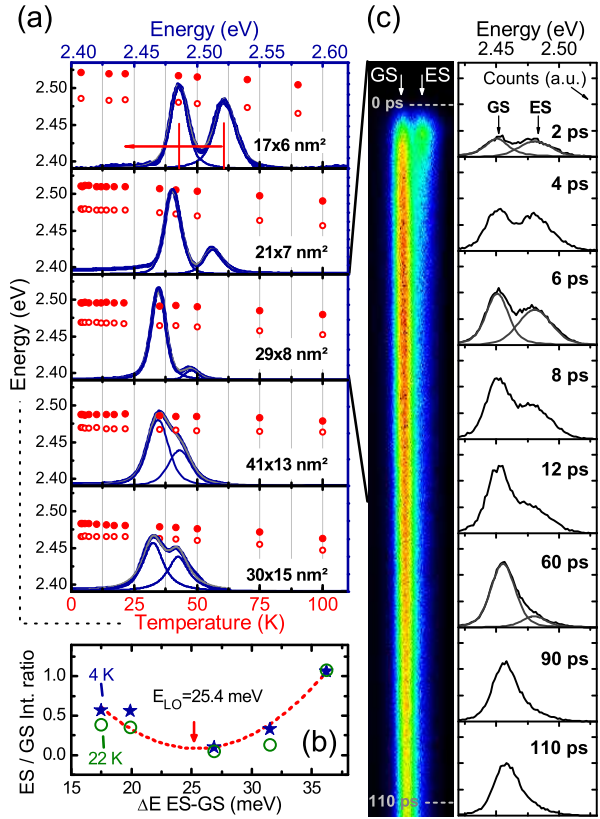}
\caption{\label{fig:spectra} (a) Time-integrated PL emission of  4.5\, monolayer (ML) CdSe NPLs with increasing lateral size from top to bottom at 4\,K (blue lines). The ground-state (GS) and excited-state (ES) emission peaks are fitted with Voigt-profiles. The resulting peak positions are plotted as open (GS) and solid (ES) red dots vs. temperature. (b) Phonon Bottleneck: ES--GS integrated PL intensity ratios  deduced from (a) vs. the ES--GS energy spacing at 4\,K and 22\,K. (c) Transient PL decay and evolution of the ES and GS emission with time and exemplary Voigt fits (platelet size 29x8\,nm$^2$).}
\end{figure}

 The PL spectra of the  investigated CdSe core NPLs are displayed in Fig.~\ref{fig:spectra} (a) showing a dual emission. Voigt-fits are used to determine the peak centers of the ground and excited state emission plotted vs. temperature as open (GS) and closed (ES) red dots. The resulting integrated peak area ratios of ES and GS are plotted in 
Fig.~\ref{fig:spectra} (b) versus the ES--GS energy difference $\Delta E$ showing a distinct minimum to be discussed later. The ES--GS energy differences $\Delta E$ are also presented in Figure 4 (a) as blue open circles. It can be seen that the energy difference increases from 18 to 36\,meV with increasing lateral platelet confinement (a table of the experimental results is presented in the Supporting Material). We conclude further that this energy spacing observed in PL can not be related to an LO phonon replica, which would have a nearly confinement independent energy spacing to the zero phonon emission.

To assess the luminescence dynamics of both PL emissions we apply time-resolved PL using a streak camera in two different time scales. Fig. 2 (c) displays a PL transient of 29x8\,nm$^2$ platelets along with spectral cuts in time showing the evolution of the dual ES--GS emission. A fast ES PL decay and a slower GS PL decay can be observed and separated by fitting the spectral contributions vs. time (Figure 3 (a) lower part).
\begin{figure}[t]
\includegraphics{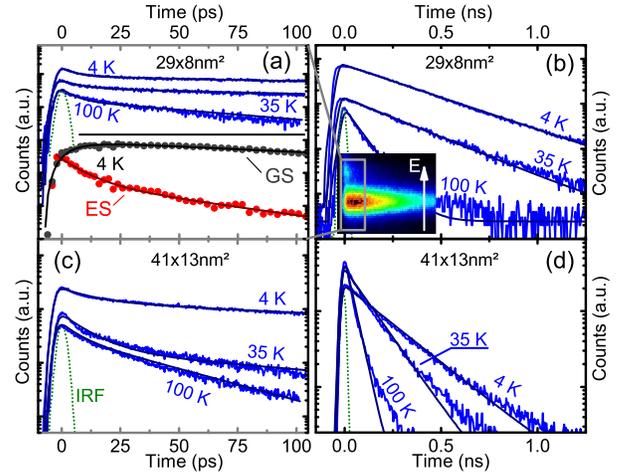}
\caption{\label{fig:transients} Blue lines: Representative examples of photo-luminescence decay curves at 4, 35 and 100\,K of CdSe NPLs with lateral size of 29x8\,nm$^2$ (panels a and b) and 41x13\,nm$^2$ (panels c and d). The bi-exponential fits (on top of data) to the fast time-range decays (a and c) use the long decay time derived from mono-exponential fits to the curves in (b) and (d) recorded in a wider time window. The instrument response function (green line) is used for deconvolution . (a) lower panel: ES (red) and GS (black) transients obtained from Voigt fits in Fig. 2 (c). Inset: Spectrally dispersed Streak Camera image of the PL decay in the first $0.5$\,ns of 29x8\,nm$^2$ CdSe NPLs at $4$\,K. Excited state emission is clearly visible. The time range of panel (a) is indicated by a grey frame.}
\end{figure}

\begin{figure}[t]
\includegraphics{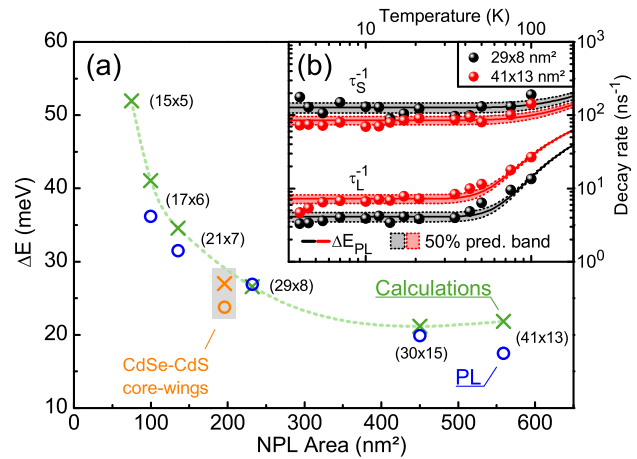}
\caption{\label{fig:deltae} (a) Energy difference $\Delta E$ between the hh ground-state and excited-state luminescence from time integrated PL (Fig.\,\ref{fig:spectra}) and our calculations plotted vs. the NPLs' CdSe core area (lateral extensions in nm (in parenthesis), dotted line only a guide to the eye). CdSe-CdS core wings NPLs (8x8\,nm$^2$ core size) were also measured (Supp. Material). (b) Decay rates of 29x8\,nm$^2$ and 41x13\,nm$^2$ NPLs from time-resolved PL measurements vs. temperature. The PL energy difference $\Delta E^{PL}$ was held fixed when fitting Eqs.\,(1) and (2) to the fast and slow decay rates, $\tau^{-1}_S$ and $\tau^{-1}_L$ (solid lines). The 50\,\% prediction bands of the fits are shown as dotted lines.}
\end{figure}

 Fig.~\ref{fig:transients} (a) and (b) show the obtained PL decay transients for 29x8\,nm$^2$ 4.5\,ML NPLs in the fast (a) and slow (b) time range at different temperatures, (c) and (d) for 41x13\,nm$^2$ platelets. To avoid fitting ambiguities the PL transients in the fast time range are fitted with bi-exponential PL decays ($I=A_1\cdot \text{exp}(t/\tau_L)+A_2\cdot \text{exp}(t/\tau_S)$) using the long time constant $\tau_L$ obtained from a mono-exponential fit to the long time range PL decay -- panels (b) and (d) -- as fixed parameter. In the long time range $\tau_L$ ($\sim 0.30$\,ns (29x8\,nm$^2$) and 0.22\,ns (41x13\,nm$^2$) at 4\,K) clearly dominates the long decay dynamics. With increasing lateral platelet size the PL decay tends to become faster in line with the predictions of the Giant Oscillator Strength (GOST) effect \cite{Feldmann1987}. 
The lower panel of Figure 3 (a) shows the decay of the ES and GS obtained from Voigt fits in Figure 2 (c) for 2\,ps temporal bins at 4\,K. It can be seen that the ES (red) fills the GS (black). A fit including the instrument response results in a 12\,ps GS filling time scale. The ground state decays then with a much longer 220\,ps time constant.
The inset in Fig.\,\ref{fig:transients}\,(b) clearly shows the dual PL emission of the ground and excited state, with an ultra fast ES recombination as in Fig. 3 (a). The results of temperature dependent PL decay fits for the samples shown in Fig.\,3 are plotted as inverse decay times in the inset (b) of Fig.\,4. \cite{BiaCom} To compare the experimentally obtained ES--GS energy spacings $\Delta E$ from time-integrated PL with time resolved measurements, we treat the ES, GS and vacuum state level as a three level system with a phonon mediated, thermally dependent, scattering channel (rate $\gamma_0$) between ES and GS analogous to Refs.\,\citenum{Labeau2003, Biadala2014a}, to investigate a phonon bottleneck. This system features a bi-exponential PL decay with a short ($\tau_{S}$) and a long decay constant ($\tau_{L}$), and intrinsic ES and GS decay rates $\Gamma_{ES}$ and $\Gamma_{GS}$:
\begin{align}
\tau_{S}^{-1}=\gamma_0(1+2/[\text{exp}(\Delta E/k_BT)-1])\\
\tau_{L}^{-1}=\frac{\Gamma_{ES}+\Gamma_{GS}}{2}-\left(\frac{\Gamma_{ES}-\Gamma_{GS}}{2}\right)\text{tanh}\left[\frac{\Delta E}{2k_B T}\right] 
\label{eq:tau}
\end{align}
Using the results of bi-exponential PL decay fits (Fig.\,3 (a) and (c)) the plotted corresponding fast ($\tau_S^{-1}$) and slow ($\tau_L^{-1}$) inverse decay constants in Fig.\,4\,(b) are fitted with Eqs.\,(1) and (2) versus temperature. In order to reduce fitting ambiguities related to the four partially correlated parameters, the corresponding experimental $\Delta E^{PL}$ value in Fig.\,\ref{fig:deltae}\,(a) is used as fixed parameter. These energy differences $\Delta E^{PL}$ -- derived from time integrated PL -- are confirmed by the plotted 50\% prediction bands for both the long and the short component of the biexponential PL decay. This is an independent confirmation that we really observe a ES--GS energy spacing in PL. Further, it turns out that the long component ($\tau_L$) is more sensitive to temperature changes than the short one and hence provides the more conclusive argument for the ES--GS energy spacing. The applicability of the inferred three level model lets us conclude that the bimodal PL distribution is related to a phonon bottleneck between ES and GS and a very large oscillator strength of both GS and ES to vacuum state transitions enabling to observe both emissions in PL at the same time. 

The observed increase of the ES--GS energy spacing with confinement (Fig. 4\,(a))  is expected for a strongly confined system. \cite{Norris1996,Moreels2011} 
Further, e.g. for the 29x8\,nm$^2$ NPLs the low temperature scattering rate $\gamma_0$ ($\tau_S^{-1}=\gamma_0=8\cdot 10^{10}\,\text{s}^{-1} \text{ for } T\rightarrow 0\,$K) of the rate equation model, representing the phonon bottleneck, is in good agreement with the fast rise kinetics of the GS emission ($12\cdot 10^{10}$\,\text{s}$^{-1}$) derived from Fig.\,3\,(a). This is an additional confirmation of the existence of a phonon bottleneck dominating the ES--GS dynamics in CdSe NPLs. The strong transversal confinement related suppression of energetically matching LO phonon modes in the NPLs \cite{Achtstein2012}, also observed as lifetime limited dephasing rates \cite{Naeem2015} results here in a slowdown of the ES--GS exciton transfer rate.
For our platelets of different lateral size this LO phonon bottleneck is also observed in the ratios of the time integrated ES--GS emissions when correlated to the corresponding ES--GS energy spacings in Figure 2\,(b). A clear minimum can be seen as the spacing approaches the CdSe platelet LO phonon energy of 25.4\,meV \cite{Cherevkov2013} (plotted for exemplary temperatures of 4 and 22\,K). In case of resonance the high density of LO phonons energetically matching the ES--GS spacing leads to a fast relaxation of the ES population to the GS resulting in a small ES--GS intensity ratio. In the off resonant cases for low and high ES-GS spacings the ES$\rightarrow$ GS relaxation is more suppressed leading to a higher ES contribution to the emission. This is again a clear manifestation of a phonon bottleneck in our NPLs.
In the following we will compare our experimental ES-GS energy differences with theoretical calculations. 

\begin{figure}[h]
\includegraphics{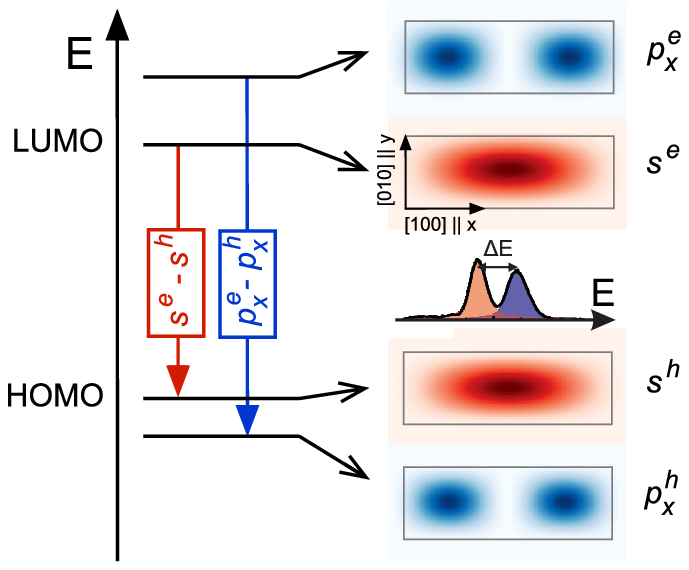}
\caption{\label{fig:theory}  Overview of the allowed optical transitions observed in experiment along with wave-function probability density plots in real space and an observed ES--GS PL spectrum.}
\end{figure}

The samples considered in this work, have different lateral sizes in $l_x$ and $l_y$ direction and a much smaller thickness $l_z$. For such structures, excited states (p-states), energetically well separated from the ground state, are expected. Figure~1 shows the evolution of the electron p-shell as the level of structural anisotropy increases up to $l_x > l_y > l_z$. In this case the $p_x$ state constitutes the lowest excited state followed by the $p_y$ and the $p_z$ state. As the vertical platelet dimension is very small, the $p_z$ state is expected at very high energies. A similar reasoning applies to the excited hole states. Fig.~\ref{fig:theory} provides an overview of the lowest allowed optical transitions: The ground state transition occurs between the electron and hole $s$-states, the first excited state transition is related to the lowest $p$-states, which we label $p_x$, as the $x$-axis corresponds to the longest axis of the platelet. 

The electronic properties and optical transitions of our 4.5\,ML CdSe NPLs are calculated in accordance with TEM data (Supporting Material). Following our previous work \cite{Achtstein2012} the electronic structure is obtained using a 3D implementation of eight-band $\mathbf{k}\!\cdot\!\mathbf{p}$ envelope function theory.
The Coulomb interaction is taken into account via a Hartree self-consistency cycle, performed separately for electron and hole ground state and their first excited states. Both, the effects arising from the dielectric environment and the electron and hole self-energy are included. 
The calculated heavy hole transition energies are compared to experimental values in table 1 (Supplementary Material): $E_{GS}^{(Theo)}$ corresponds to the calculated energy of the ground state exciton, and $\Delta E^{(Theo)}$ is the energy difference to the excited state transition, $p_{x}^{e} - p_{x}^{h}$ shown in Figure 4 (a). The theoretical and experimental values of both quantities show a very good agreement (see also supp. Material), thus, supporting the direct observation of an excited state - ground state energy spacing $\Delta E$ in our experiments. 

In summary we have shown that CdSe NPLs exhibit not only lowest hh s-exciton state (GS) related photoluminescence upon continuum excitation, but also p-state (ES) luminescence far below GS saturation. Calculations and time integrated PL show a strong increase of the ES--GS energy spacing from about 18 to 36\,meV with increasing quantization of the exciton wavefunction. The existence of a phonon bottleneck between ES and GS is confirmed by three methods: A rate equation model for the temperature dependence, the temporal course of ES and GS emission and the observation of an ES/GS intensity ratio minimum in the time-integrated PL for size dependent ES--GS energy spacings resonant to the LO phonon energy of 25.4\,meV in the CdSe NPLs. We conclude further that the presented double emission in PL is not related to an LO phonon replica, which would have the same GS dynamics and a practically confinement independent energy spacing to the GS emission. In contrast the observed bi-exponential PL decay of nanoplatelets is related to a phonon bottleneck between ES and GS populations. \\
\begin{acknowledgments}
AWA, RS and UW  acknowledges DFG Projects AC290 and WO477, MA the CHEMREAGENTS program.
\end{acknowledgments}

\bibliography{apssamp}

\begin{thebibliography}{16}%
\makeatletter
\providecommand \@ifxundefined [1]{%
 \@ifx{#1\undefined}
}%
\providecommand \@ifnum [1]{%
 \ifnum #1\expandafter \@firstoftwo
 \else \expandafter \@secondoftwo
 \fi
}%
\providecommand \@ifx [1]{%
 \ifx #1\expandafter \@firstoftwo
 \else \expandafter \@secondoftwo
 \fi
}%
\providecommand \natexlab [1]{#1}%
\providecommand \enquote  [1]{``#1''}%
\providecommand \bibnamefont  [1]{#1}%
\providecommand \bibfnamefont [1]{#1}%
\providecommand \citenamefont [1]{#1}%
\providecommand \href@noop [0]{\@secondoftwo}%
\providecommand \href [0]{\begingroup \@sanitize@url \@href}%
\providecommand \@href[1]{\@@startlink{#1}\@@href}%
\providecommand \@@href[1]{\endgroup#1\@@endlink}%
\providecommand \@sanitize@url [0]{\catcode `\\12\catcode `\$12\catcode
  `\&12\catcode `\#12\catcode `\^12\catcode `\_12\catcode `\%12\relax}%
\providecommand \@@startlink[1]{}%
\providecommand \@@endlink[0]{}%
\providecommand \url  [0]{\begingroup\@sanitize@url \@url }%
\providecommand \@url [1]{\endgroup\@href {#1}{\urlprefix }}%
\providecommand \urlprefix  [0]{URL }%
\providecommand \Eprint [0]{\href }%
\providecommand \doibase [0]{http://dx.doi.org/}%
\providecommand \selectlanguage [0]{\@gobble}%
\providecommand \bibinfo  [0]{\@secondoftwo}%
\providecommand \bibfield  [0]{\@secondoftwo}%
\providecommand \translation [1]{[#1]}%
\providecommand \BibitemOpen [0]{}%
\providecommand \bibitemStop [0]{}%
\providecommand \bibitemNoStop [0]{.\EOS\space}%
\providecommand \EOS [0]{\spacefactor3000\relax}%
\providecommand \BibitemShut  [1]{\csname bibitem#1\endcsname}%
\let\auto@bib@innerbib\@empty
\bibitem [{\citenamefont {Antanovich}\ \emph {et~al.}(2015)\citenamefont
  {Antanovich}, \citenamefont {Prudnikau}, \citenamefont {Melnikau},
  \citenamefont {Rakovich}, \citenamefont {Chuvilin}, \citenamefont {Woggon},
  \citenamefont {Achtstein},\ and\ \citenamefont {Artemyev}}]{Antan2015}%
  \BibitemOpen
  \bibfield  {author} {\bibinfo {author} {\bibfnamefont {A.~V.}\ \bibnamefont
  {Antanovich}}, \bibinfo {author} {\bibfnamefont {A.~V.}\ \bibnamefont
  {Prudnikau}}, \bibinfo {author} {\bibfnamefont {D.}~\bibnamefont {Melnikau}},
  \bibinfo {author} {\bibfnamefont {Y.~P.}\ \bibnamefont {Rakovich}}, \bibinfo
  {author} {\bibfnamefont {A.}~\bibnamefont {Chuvilin}}, \bibinfo {author}
  {\bibfnamefont {U.}~\bibnamefont {Woggon}}, \bibinfo {author} {\bibfnamefont
  {A.~W.}\ \bibnamefont {Achtstein}}, \ and\ \bibinfo {author} {\bibfnamefont
  {M.~V.}\ \bibnamefont {Artemyev}},\ }\href {\doibase 10.1039/C4NR07134D}
  {\bibfield  {journal} {\bibinfo  {journal} {Nanoscale}\ }\textbf {\bibinfo
  {volume} {7}},\ \bibinfo {pages} {8084} (\bibinfo {year} {2015})}\BibitemShut
  {NoStop}%
\bibitem [{\citenamefont {Ithurria}\ \emph {et~al.}(2011)\citenamefont
  {Ithurria}, \citenamefont {Tessier}, \citenamefont {Mahler}, \citenamefont
  {Lobo}, \citenamefont {Dubertret},\ and\ \citenamefont
  {Efros}}]{Ithurria2011a}%
  \BibitemOpen
  \bibfield  {author} {\bibinfo {author} {\bibfnamefont {S.}~\bibnamefont
  {Ithurria}}, \bibinfo {author} {\bibfnamefont {M.~D.}\ \bibnamefont
  {Tessier}}, \bibinfo {author} {\bibfnamefont {B.}~\bibnamefont {Mahler}},
  \bibinfo {author} {\bibfnamefont {R.~P. S.~M.}\ \bibnamefont {Lobo}},
  \bibinfo {author} {\bibfnamefont {B.}~\bibnamefont {Dubertret}}, \ and\
  \bibinfo {author} {\bibfnamefont {A.~L.}\ \bibnamefont {Efros}},\ }\href
  {http://dx.doi.org/10.1038/nmat3145} {\bibfield  {journal} {\bibinfo
  {journal} {Nat. Mater.}\ }\textbf {\bibinfo {volume} {10}},\ \bibinfo {pages}
  {936} (\bibinfo {year} {2011})}\BibitemShut {NoStop}%
\bibitem [{\citenamefont {Achtstein}\ \emph {et~al.}(2012)\citenamefont
  {Achtstein}, \citenamefont {Schliwa}, \citenamefont {Prudnikau},
  \citenamefont {Hardzei}, \citenamefont {Artemyev}, \citenamefont {Thomsen},\
  and\ \citenamefont {Woggon}}]{Achtstein2012}%
  \BibitemOpen
  \bibfield  {author} {\bibinfo {author} {\bibfnamefont {A.~W.}\ \bibnamefont
  {Achtstein}}, \bibinfo {author} {\bibfnamefont {A.}~\bibnamefont {Schliwa}},
  \bibinfo {author} {\bibfnamefont {A.}~\bibnamefont {Prudnikau}}, \bibinfo
  {author} {\bibfnamefont {M.}~\bibnamefont {Hardzei}}, \bibinfo {author}
  {\bibfnamefont {M.~V.}\ \bibnamefont {Artemyev}}, \bibinfo {author}
  {\bibfnamefont {C.}~\bibnamefont {Thomsen}}, \ and\ \bibinfo {author}
  {\bibfnamefont {U.}~\bibnamefont {Woggon}},\ }\href {\doibase
  10.1021/nl301071n} {\bibfield  {journal} {\bibinfo  {journal} {Nano Lett.}\
  }\textbf {\bibinfo {volume} {12}},\ \bibinfo {pages} {3151} (\bibinfo {year}
  {2012})}\BibitemShut {NoStop}%
\bibitem [{\citenamefont {Naeem}\ \emph {et~al.}(2015)\citenamefont {Naeem},
  \citenamefont {Masia}, \citenamefont {Christodoulou}, \citenamefont
  {Moreels}, \citenamefont {Borri},\ and\ \citenamefont
  {Langbein}}]{Naeem2015}%
  \BibitemOpen
  \bibfield  {author} {\bibinfo {author} {\bibfnamefont {A.}~\bibnamefont
  {Naeem}}, \bibinfo {author} {\bibfnamefont {F.}~\bibnamefont {Masia}},
  \bibinfo {author} {\bibfnamefont {S.}~\bibnamefont {Christodoulou}}, \bibinfo
  {author} {\bibfnamefont {I.}~\bibnamefont {Moreels}}, \bibinfo {author}
  {\bibfnamefont {P.}~\bibnamefont {Borri}}, \ and\ \bibinfo {author}
  {\bibfnamefont {W.}~\bibnamefont {Langbein}},\ }\href {\doibase
  10.1103/PhysRevB.91.121302} {\bibfield  {journal} {\bibinfo  {journal} {Phys.
  Rev. B}\ }\textbf {\bibinfo {volume} {91}},\ \bibinfo {pages} {121302}
  (\bibinfo {year} {2015})}\BibitemShut {NoStop}%
\bibitem [{\citenamefont {Cassette}\ \emph {et~al.}(2015)\citenamefont
  {Cassette}, \citenamefont {Pensack}, \citenamefont {Mahler},\ and\
  \citenamefont {Scholes}}]{Cassette2015}%
  \BibitemOpen
  \bibfield  {author} {\bibinfo {author} {\bibfnamefont {E.}~\bibnamefont
  {Cassette}}, \bibinfo {author} {\bibfnamefont {R.~D.}\ \bibnamefont
  {Pensack}}, \bibinfo {author} {\bibfnamefont {B.}~\bibnamefont {Mahler}}, \
  and\ \bibinfo {author} {\bibfnamefont {G.~D.}\ \bibnamefont {Scholes}},\
  }\href {http://dx.doi.org/10.1038/ncomms7086} {\bibfield  {journal} {\bibinfo
   {journal} {Nat. Commun.}\ }\textbf {\bibinfo {volume} {6}},\  (\bibinfo
  {year} {2015})}\BibitemShut {NoStop}%
\bibitem [{\citenamefont {Grim}\ \emph {et~al.}(2014)\citenamefont {Grim},
  \citenamefont {Christodoulou}, \citenamefont {{Di Stasio}}, \citenamefont
  {Krahne}, \citenamefont {Cingolani}, \citenamefont {Manna},\ and\
  \citenamefont {Moreels}}]{Grim2014}%
  \BibitemOpen
  \bibfield  {author} {\bibinfo {author} {\bibfnamefont {J.~Q.}\ \bibnamefont
  {Grim}}, \bibinfo {author} {\bibfnamefont {S.}~\bibnamefont {Christodoulou}},
  \bibinfo {author} {\bibfnamefont {F.}~\bibnamefont {{Di Stasio}}}, \bibinfo
  {author} {\bibfnamefont {R.}~\bibnamefont {Krahne}}, \bibinfo {author}
  {\bibfnamefont {R.}~\bibnamefont {Cingolani}}, \bibinfo {author}
  {\bibfnamefont {L.}~\bibnamefont {Manna}}, \ and\ \bibinfo {author}
  {\bibfnamefont {I.}~\bibnamefont {Moreels}},\ }\href {\doibase
  10.1038/nnano.2014.213} {\bibfield  {journal} {\bibinfo  {journal} {Nature
  Nanotechnology}\ }\textbf {\bibinfo {volume} {9}},\ \bibinfo {pages} {891}
  (\bibinfo {year} {2014})}\BibitemShut {NoStop}%
\bibitem [{\citenamefont {Achtstein}\ \emph {et~al.}(2014)\citenamefont
  {Achtstein}, \citenamefont {Prudnikau}, \citenamefont {Ermolenko},
  \citenamefont {Gurinovich}, \citenamefont {Gaponenko}, \citenamefont
  {Woggon}, \citenamefont {Baranov}, \citenamefont {Leonov}, \citenamefont
  {Rukhlenko}, \citenamefont {Fedorov},\ and\ \citenamefont
  {Artemyev}}]{Achtstein2014}%
  \BibitemOpen
  \bibfield  {author} {\bibinfo {author} {\bibfnamefont {A.~W.}\ \bibnamefont
  {Achtstein}}, \bibinfo {author} {\bibfnamefont {A.~V.}\ \bibnamefont
  {Prudnikau}}, \bibinfo {author} {\bibfnamefont {M.~V.}\ \bibnamefont
  {Ermolenko}}, \bibinfo {author} {\bibfnamefont {L.~I.}\ \bibnamefont
  {Gurinovich}}, \bibinfo {author} {\bibfnamefont {S.~V.}\ \bibnamefont
  {Gaponenko}}, \bibinfo {author} {\bibfnamefont {U.}~\bibnamefont {Woggon}},
  \bibinfo {author} {\bibfnamefont {A.~V.}\ \bibnamefont {Baranov}}, \bibinfo
  {author} {\bibfnamefont {M.~Y.}\ \bibnamefont {Leonov}}, \bibinfo {author}
  {\bibfnamefont {I.~D.}\ \bibnamefont {Rukhlenko}}, \bibinfo {author}
  {\bibfnamefont {A.~V.}\ \bibnamefont {Fedorov}}, \ and\ \bibinfo {author}
  {\bibfnamefont {M.~V.}\ \bibnamefont {Artemyev}},\ }\href {\doibase
  10.1021/nn503745u} {\bibfield  {journal} {\bibinfo  {journal} {ACS Nano}\
  }\textbf {\bibinfo {volume} {8}},\ \bibinfo {pages} {7678} (\bibinfo {year}
  {2014})}\BibitemShut {NoStop}%
\bibitem [{\citenamefont {Biadala}\ \emph
  {et~al.}(2014{\natexlab{a}})\citenamefont {Biadala}, \citenamefont {Liu},
  \citenamefont {Tessier}, \citenamefont {Yakovlev}, \citenamefont
  {Dubertret},\ and\ \citenamefont {Bayer}}]{Biadala2014}%
  \BibitemOpen
  \bibfield  {author} {\bibinfo {author} {\bibfnamefont {L.}~\bibnamefont
  {Biadala}}, \bibinfo {author} {\bibfnamefont {F.}~\bibnamefont {Liu}},
  \bibinfo {author} {\bibfnamefont {M.~D.}\ \bibnamefont {Tessier}}, \bibinfo
  {author} {\bibfnamefont {D.~R.}\ \bibnamefont {Yakovlev}}, \bibinfo {author}
  {\bibfnamefont {B.}~\bibnamefont {Dubertret}}, \ and\ \bibinfo {author}
  {\bibfnamefont {M.}~\bibnamefont {Bayer}},\ }\href {\doibase
  10.1021/nl403311n} {\bibfield  {journal} {\bibinfo  {journal} {Nano Lett.}\
  }\textbf {\bibinfo {volume} {14}},\ \bibinfo {pages} {1134} (\bibinfo {year}
  {2014}{\natexlab{a}})}\BibitemShut {NoStop}%
\bibitem [{\citenamefont {She}\ \emph {et~al.}(2014)\citenamefont {She},
  \citenamefont {Fedin}, \citenamefont {Dolzhnikov}, \citenamefont
  {Demorti{\`e}re}, \citenamefont {Schaller}, \citenamefont {Pelton},\ and\
  \citenamefont {Talapin}}]{She2014}%
  \BibitemOpen
  \bibfield  {author} {\bibinfo {author} {\bibfnamefont {C.}~\bibnamefont
  {She}}, \bibinfo {author} {\bibfnamefont {I.}~\bibnamefont {Fedin}}, \bibinfo
  {author} {\bibfnamefont {D.~S.}\ \bibnamefont {Dolzhnikov}}, \bibinfo
  {author} {\bibfnamefont {A.}~\bibnamefont {Demorti{\`e}re}}, \bibinfo
  {author} {\bibfnamefont {R.~D.}\ \bibnamefont {Schaller}}, \bibinfo {author}
  {\bibfnamefont {M.}~\bibnamefont {Pelton}}, \ and\ \bibinfo {author}
  {\bibfnamefont {D.~V.}\ \bibnamefont {Talapin}},\ }\href@noop {} {\bibfield
  {journal} {\bibinfo  {journal} {Nano Lett.}\ }\textbf {\bibinfo {volume}
  {14}},\ \bibinfo {pages} {2772} (\bibinfo {year} {2014})}\BibitemShut
  {NoStop}%
\bibitem [{\citenamefont {Feldmann}\ \emph {et~al.}(1987)\citenamefont
  {Feldmann}, \citenamefont {Peter}, \citenamefont {G\"obel}, \citenamefont
  {Dawson}, \citenamefont {Moore}, \citenamefont {Foxon},\ and\ \citenamefont
  {Elliott}}]{Feldmann1987}%
  \BibitemOpen
  \bibfield  {author} {\bibinfo {author} {\bibfnamefont {J.}~\bibnamefont
  {Feldmann}}, \bibinfo {author} {\bibfnamefont {G.}~\bibnamefont {Peter}},
  \bibinfo {author} {\bibfnamefont {E.~O.}\ \bibnamefont {G\"obel}}, \bibinfo
  {author} {\bibfnamefont {P.}~\bibnamefont {Dawson}}, \bibinfo {author}
  {\bibfnamefont {K.}~\bibnamefont {Moore}}, \bibinfo {author} {\bibfnamefont
  {C.}~\bibnamefont {Foxon}}, \ and\ \bibinfo {author} {\bibfnamefont {R.~J.}\
  \bibnamefont {Elliott}},\ }\href {\doibase 10.1103/PhysRevLett.59.2337}
  {\bibfield  {journal} {\bibinfo  {journal} {Phys. Rev. Lett.}\ }\textbf
  {\bibinfo {volume} {59}},\ \bibinfo {pages} {2337} (\bibinfo {year}
  {1987})}\BibitemShut {NoStop}%
\bibitem [{Bia()}]{BiaCom}%
  \BibitemOpen
  \href@noop {} {}\bibinfo {note} {In line with observations of Biadala et al.
  \cite{Biadala2014} we observe at very low temperatures a slight bending in
  the course of the long component ($\tau_L^{-1}$), which may be related to a
  $\sim$ meV fine structure splitting of the GS.}\BibitemShut {Stop}%
\bibitem [{\citenamefont {Labeau}\ \emph {et~al.}(2003)\citenamefont {Labeau},
  \citenamefont {Tamarat},\ and\ \citenamefont {Lounis}}]{Labeau2003}%
  \BibitemOpen
  \bibfield  {author} {\bibinfo {author} {\bibfnamefont {O.}~\bibnamefont
  {Labeau}}, \bibinfo {author} {\bibfnamefont {P.}~\bibnamefont {Tamarat}}, \
  and\ \bibinfo {author} {\bibfnamefont {B.}~\bibnamefont {Lounis}},\ }\href
  {\doibase 10.1103/PhysRevLett.90.257404} {\bibfield  {journal} {\bibinfo
  {journal} {Phys. Rev. Lett.}\ }\textbf {\bibinfo {volume} {90}},\ \bibinfo
  {pages} {257404} (\bibinfo {year} {2003})}\BibitemShut {NoStop}%
\bibitem [{\citenamefont {Biadala}\ \emph
  {et~al.}(2014{\natexlab{b}})\citenamefont {Biadala}, \citenamefont {Siebers},
  \citenamefont {Gomes}, \citenamefont {Hens}, \citenamefont {Yakovlev},\ and\
  \citenamefont {Bayer}}]{Biadala2014a}%
  \BibitemOpen
  \bibfield  {author} {\bibinfo {author} {\bibfnamefont {L.}~\bibnamefont
  {Biadala}}, \bibinfo {author} {\bibfnamefont {B.}~\bibnamefont {Siebers}},
  \bibinfo {author} {\bibfnamefont {R.}~\bibnamefont {Gomes}}, \bibinfo
  {author} {\bibfnamefont {Z.}~\bibnamefont {Hens}}, \bibinfo {author}
  {\bibfnamefont {D.~R.}\ \bibnamefont {Yakovlev}}, \ and\ \bibinfo {author}
  {\bibfnamefont {M.}~\bibnamefont {Bayer}},\ }\href {\doibase
  10.1021/jp505887u} {\bibfield  {journal} {\bibinfo  {journal} {J. Phys. Chem.
  C}\ }\textbf {\bibinfo {volume} {118}},\ \bibinfo {pages} {22309} (\bibinfo
  {year} {2014}{\natexlab{b}})}\BibitemShut {NoStop}%
\bibitem [{\citenamefont {Norris}\ and\ \citenamefont
  {Bawendi}(1996)}]{Norris1996}%
  \BibitemOpen
  \bibfield  {author} {\bibinfo {author} {\bibfnamefont {D.~J.}\ \bibnamefont
  {Norris}}\ and\ \bibinfo {author} {\bibfnamefont {M.~G.}\ \bibnamefont
  {Bawendi}},\ }\href {\doibase 10.1103/PhysRevB.53.16338} {\bibfield
  {journal} {\bibinfo  {journal} {Phys. Rev. B}\ }\textbf {\bibinfo {volume}
  {53}},\ \bibinfo {pages} {16338} (\bibinfo {year} {1996})}\BibitemShut
  {NoStop}%
\bibitem [{\citenamefont {Moreels}\ \emph {et~al.}(2011)\citenamefont
  {Moreels}, \citenamefont {Rainò}, \citenamefont {Gomes}, \citenamefont
  {Hens}, \citenamefont {Stöferle},\ and\ \citenamefont {Mahrt}}]{Moreels2011}%
  \BibitemOpen
  \bibfield  {author} {\bibinfo {author} {\bibfnamefont {I.}~\bibnamefont
  {Moreels}}, \bibinfo {author} {\bibfnamefont {G.}~\bibnamefont {Rainò}},
  \bibinfo {author} {\bibfnamefont {R.}~\bibnamefont {Gomes}}, \bibinfo
  {author} {\bibfnamefont {Z.}~\bibnamefont {Hens}}, \bibinfo {author}
  {\bibfnamefont {T.}~\bibnamefont {Stöferle}}, \ and\ \bibinfo {author}
  {\bibfnamefont {R.~F.}\ \bibnamefont {Mahrt}},\ }\href {\doibase
  10.1021/nn202604z} {\bibfield  {journal} {\bibinfo  {journal} {ACS Nano}\
  }\textbf {\bibinfo {volume} {5}},\ \bibinfo {pages} {8033} (\bibinfo {year}
  {2011})}\BibitemShut {NoStop}%
\bibitem [{\citenamefont {Cherevkov}\ \emph {et~al.}(2013)\citenamefont
  {Cherevkov}, \citenamefont {Artemyev}, \citenamefont {Prudnikau},\ and\
  \citenamefont {Baranov}}]{Cherevkov2013}%
  \BibitemOpen
  \bibfield  {author} {\bibinfo {author} {\bibfnamefont {S.~A.}\ \bibnamefont
  {Cherevkov}}, \bibinfo {author} {\bibfnamefont {M.~V.}\ \bibnamefont
  {Artemyev}}, \bibinfo {author} {\bibfnamefont {A.~V.}\ \bibnamefont
  {Prudnikau}}, \ and\ \bibinfo {author} {\bibfnamefont {A.~V.}\ \bibnamefont
  {Baranov}},\ }\href@noop {} {\bibfield  {journal} {\bibinfo  {journal} {Phys.
  Rev. B}\ }\textbf {\bibinfo {volume} {88}},\ \bibinfo {pages} {041303}
  (\bibinfo {year} {2013})}\BibitemShut {NoStop}%
\end{thebibliography}%

\end{document}